\documentclass[twocolumn]{aastex631}
\usepackage{xspace}
\usepackage{subfigure}
\usepackage{booktabs}
\usepackage{longtable}
\usepackage{multirow}
\usepackage{rotating}
\usepackage{tabularx}
\usepackage{appendix}
\usepackage{amsmath}
\usepackage{xcolor}
\usepackage{placeins}

\begin{document}

\title{JWST's PEARLS: A clumpy ring galaxy at $z = 4.0148$}

\author[0000-0001-7610-5544]{David Vizgan}
\affiliation{Department of Astronomy, University of Illinois, 1002 West Green Street, Urbana, IL, 61801, USA}

\author[0000-0001-5105-2837]{Ming-Yang Zhuang}
\affiliation{Department of Astronomy, University of Illinois, 1002 West Green Street, Urbana, IL, 61801, USA}

\author[0000-0003-3037-257X]{Ian Smail}
\affiliation{Centre for Extragalactic Astronomy, Department of Physics, Durham University, South Road, Durham DH1, 3LE, UK}

\author[0000-0001-8156-6281]{Rogier A. Windhorst}
\affiliation{School of Earth and Space Exploration, Arizona State University,
Tempe, AZ 85287-6004, USA}

\author[0009-0007-0782-0721]{Gibson B.\ Bowling}
\affiliation{School of Earth and Space Exploration, Arizona State University,
Tempe, AZ 85287-6004, USA}

\author[0000-0003-0202-0534]{Cheng Cheng}
\affiliation{Chinese Academy of Sciences South America Center for Astronomy, National Astronomical Observatories, CAS, Beijing 100101, China} 
\affiliation{Key Laboratory of Optical Astronomy, NAOC, 20A Datun Road, Chaoyang District, Beijing 100101, China}

\author[0000-0003-3329-1337]{Seth H.\ Cohen}
\affiliation{School of Earth and Space Exploration, Arizona State University,
Tempe, AZ 85287-6004, USA}

\author[0000-0003-1949-7638]{Christopher J. Conselice}
\affiliation{Jodrell Bank Centre for Astrophysics, University of Manchester, Oxford Road, Manchester UK}

\author[0000-0001-9065-3926]{Jose M. Diego}
\affiliation{Instituto de Fisica de Cantabria (CSIC-UC). Avenida. Los Castros s/n, E-39005 Santander, Spain}

\author[0000-0003-1625-8009]{Brenda L. Frye}
\affiliation{Department of Astronomy/Steward Observatory, University of Arizona, 933 N. Cherry Avenue, Tucson, AZ 85721, USA}

\author[0000-0001-9440-8872]{Norman A. Grogin}
\affiliation{Space Telescope Science Institute, 3700 San Martin Drive, Baltimore, MD 21218, USA}

\author[0000-0003-1268-5230]{Rolf A.\ Jansen}
\affiliation{School of Earth and Space Exploration, Arizona State University,
Tempe, AZ 85287-6004, USA}

\author[0000-0001-9394-6732]{Patrick S. Kamieneski}
\affiliation{Department of Space, Earth \& Environment, Chalmers University of Technology, 412 96 Gothenburg, Sweden}

\author[0000-0002-6610-2048]{Anton M. Koekemoer}
\affiliation{Space Telescope Science Institute, 3700 San Martin Drive, Baltimore, MD 21218, USA}

\author[0000-0002-6150-833X]{Rafael Ortiz~III} 
\affiliation{School of Earth and Space Exploration, Arizona State University,
Tempe, AZ 85287-6004, USA}

\author[0000-0003-4223-7324]{Massimo Ricotti}
\affiliation{Department of Astronomy, University of Maryland, College Park, Maryland 20742, USA}

\author[0000-0001-7957-6202]{Bangzheng Sun}
\affiliation{Department of Physics and Astronomy, University of Missouri-Columbia, 701 S College Ave, Columbia, Missouri 65201}

\author[0000-0002-1681-0767]{Hayley Williams}
\affiliation{School of Earth and Space Exploration, Arizona State University,
Tempe, AZ 85287-6004, USA}

\author[0000-0002-9895-5758]{S.\ P.\ Willner}
\affiliation{Center for Astrophysics \textbar\ Harvard \& Smithsonian, 60 Garden Street, Cambridge, MA, 02138, USA}

\author[0000-0001-7592-7714]{Haojing Yan}
\affiliation{Department of Physics and Astronomy, University of Missouri-Columbia, 701 S College Ave, Columbia, Missouri 65201}

\author[0009-0008-1965-9012]{Aadya Agrawal}
\affiliation{Department of Astronomy, University of Illinois, 1002 West Green Street, Urbana, IL, 61801, USA}

\author[0000-0001-6629-0379]{Manuel Solimano}
\affiliation{Centro de Astrobiolog\'ia (CAB), CSIC-INTA, Ctra. de Ajalvir km 4, Torrej\'on de Ardoz, E-28850, Madrid, Spain}

\author[0000-0002-8501-3518]{Zachary Stone}
\affiliation{Department of Astronomy, University of Illinois, 1002 West Green Street, Urbana, IL, 61801, USA}

\author[0000-0001-7192-3871]{Joaquin~D. Vieira}
\affiliation{Department of Astronomy, University of Illinois, 1002 West Green Street, Urbana, IL, 61801, USA}
\affiliation{Department of Physics, University of Illinois, 1110 West Green Street, Urbana, IL, 61801, USA}
\affiliation{Center for AstroPhysical Surveys, National Center for Supercomputing Applications, 1205 West Clark Street, Urbana, IL, 61820, USA}

\author[0000-0002-8117-9991]{Chentao Yang}
\affiliation{Department of Space, Earth \& Environment, Chalmers University of Technology, 412 96 Gothenburg, Sweden}

\begin{abstract}

Ring galaxies are an uncommon class of galaxies whose morphology is closely related to dynamical processes that govern galaxy evolution. Some ring galaxies, known as ``collisional ring galaxies'', are thought to form as a consequence of head-on collisions between galaxies, and a number of high-redshift collisional ring galaxies have been discovered and/or studied in the era of the James Webb Space Telescope (JWST). In this paper, we present HST/ACS, JWST/NIRCam, and JWST/NIRSpec observations of a candidate ring galaxy at $z_{\rm spec} = 4.0148$, previously identified as a potential gravitational lens. The galaxy exhibits a complex morphology, including three bright clumps along an apparent ring with radius $\approx 0.25\arcsec$ $\simeq  1.8$ kpc. It has a total SFR $= 140^{+20}_{-30}$ ${\rm M}_{\rm \odot}$ yr$^{-1}$ and $\log(M_\ast/{\rm M}_\odot) = 10.41^{+0.11}_{-0.13}$, making it similar to other high-redshift collisional ring galaxies. Although we argue strongly in favor of the collisional ring explanation, we cannot entirely rule out a galaxy-galaxy strong lensing explanation for the system's morphology, in which a foreground galaxy at $z \simeq 1.7$ lenses a galaxy at $z \simeq 4.0$ into an Einstein ring-like configuration; to confirm the nature of this source, we require kinematic information via high spectral resolution observations. We suggest that current and future gravitational lens surveys should consider high-redshift ring galaxies as possible but significant contaminants.

\end{abstract}


\section{Introduction} \label{sec:intro}

Astronomers have known for a century that galaxies exhibit stark differences in morphology \citep{hubble1926}, and for decades have studied the inextricable links between morphological features, such as spiral arms and bars, to galaxy evolution and the processes that govern it across cosmic time \citep[e.g.][]{devaucoleurs1959, kormendy2004, buta2015}. Rings are an uncommon but important morphological feature in galaxies which appear to be present in $\sim 20\%$ of spiral galaxies \citep[e.g.][]{buta1996, fernandez2021}. Most commonly, rings in galaxies are manifestations of special resonant patterns (such as inner and outer Lindblad resonances) with density waves that govern spiral structure in galaxies \citep{buta1996, buta1999, comeron2014}. 

A special case of ring galaxies is the ``collisional ring'' galaxy \citep[see][for review]{appleton1996}, which accounts for just 0.01\% of all galaxies \citep{madore2009}. The standard picture of their formation involves a galaxy collision \citep{lynds1976, theys1977, binney2008}; specifically, a smaller galaxy, often likened to a ``bullet'' in the literature \citep{parker2015, khoram2025}, collides head-on with a larger disk \citep{appleton1996}, triggering a density wave propagating outward across the disk plane. Collisional rings are predicted to be clumpy and short-lived \citep[e.g.][]{elmegreen2006, renaud2018}, and both resonant and collisional ring galaxies exhibit enhanced star-formation due to the compression of gas along wavefronts \citep[e.g.][]{appleton1987}.

Recently, ring galaxies have been identified beyond $z > 1$ \citep{yuan2020, liu2023}, particularly using JWST \citep{nestor2025, perna2025, li2025, khoram2025}. The majority of these sources are thought to have formed from collisions due to the presence of nearby companions, which have made them compelling laboratories for merger-related dynamics at high redshift. Their relative rarity in the early universe is debated; \cite{smirnov2022} analytically determined that the volume density of collisional ring galaxies should increase with cosmic time, whereas \cite{elagali2018} found in simulations that the collisional ring galaxy formation density rate should be roughly constant with redshift, a finding inferred by \cite{yuan2020} from observations. Collisional ring galaxies have been proposed as alternative signposts for galaxy mergers / merger rates across cosmic time \citep[e.g.][]{lavery2004, donghia2008}, though it has been argued with simulations that collisional ring galaxies could be a biased tracer of mergers across redshift \citep{elagali2018}. The merger and galaxy interaction rate is greater at higher redshifts \citep{duan2025}, and thus one might expect ring galaxy number densities to increase with redshift.

In this paper, we study a high-redshift galaxy near the galaxy cluster MACS J0416.1-2403, selected from the Massive Cluster Survey \citep[MACS;][]{ebeling2001}, which has been characterized as a complex merging cluster at $z = 0.397$ \citep[][]{mann2012}. Multiple works \citep[e.g.][]{jauzac2014, diego2024, rihtaric2025} have used the presence of arcs and multiply imaged galaxies, due to gravitational lensing, to build an accurate lens model for this cluster, thanks to high-resolution, deep imaging with the Hubble Space Telescope (HST) and James Webb Space Telescope (JWST), respectively. 

The galaxy examined in this work was first cataloged by \cite{merlin2016} and further discussed by \cite{diego2024}, who suggested that it was a high-redshift gravitational lens candidate. Using multi-band HST ACS and JWST NIRCam imaging, along with NIRSpec PRISM observations, we argue in this work that it is most likely a collisional ring galaxy. If this system is a natural ring galaxy, and not a galaxy-galaxy scale lens, then it is the most distant galaxy-scale ring discovered to date, with a spectroscopic redshift $z = 4.0148$. We also show in this work that this source is similar in several important aspects to other collisional ring candidates studied with JWST \citep[e.g.][]{nestor2025, perna2025}. We present multi-band imaging and spectra from HST and JWST in $\S$\ref{sec:obs}. We describe the morphology of the source in $\S$\ref{sec:res}. We study the photometry, perform spectral energy distribution (SED) fitting in $\S$\ref{sec:ana}, and contextualize this galaxy alongside similar sources in $\S$\ref{sec:disc}, concluding this work in $\S$\ref{sec:conc}. Throughout this work we assume the flat $\Lambda$CDM cosmology from the \cite{planck2020} results; $H_0 = 67.7$ km s$^{-1}$, $\Omega_0 = 0.31$, $T_{\rm cmb} = 2.725$ K. 

\begin{figure*}[ht]
    \centering
    \includegraphics[width=0.98\textwidth]{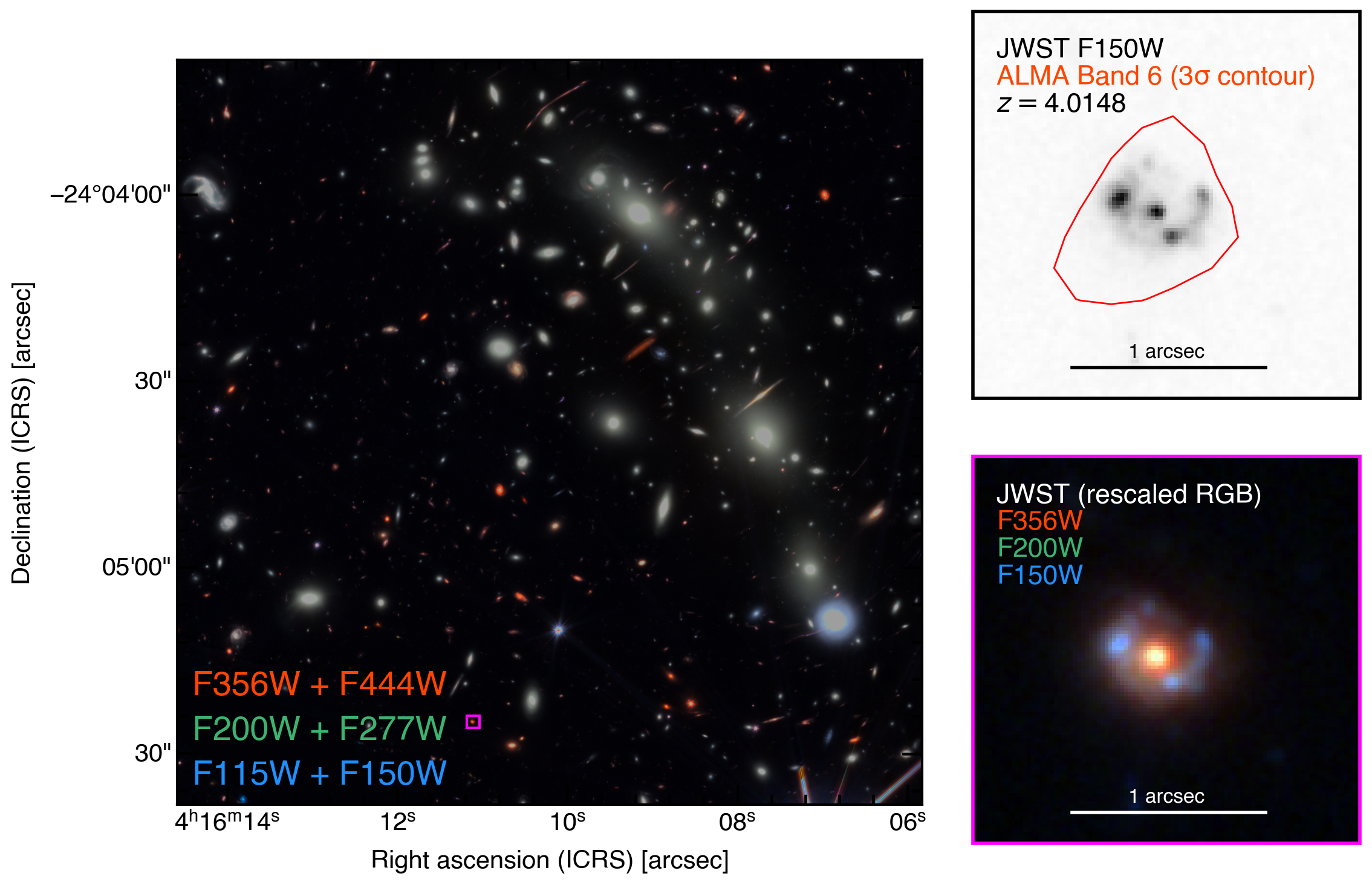}
    \caption{\textbf{Left:} 4 arcmin$^2$ pseudo-RGB image of the MACS0416 cluster, using JWST NIRCam F115W + F150W as blue, F200W + F277W as green, and F356W + F444W as red. The location of the ring galaxy is indicated in a magenta box. \textbf{Right:} $2\arcsec$ cutout of the candidate ring galaxy in JWST F150W imaging (top) and $2\arcsec$ re-scaled RGB cutout of the CRG (bottom), using the F150W, F200W, and F356W filters (corresponding to rest-frame wavelengths of 0.30, 0.40, and 0.71 $\mu$m respectively at $z = 4.0148$). In the top panel, we show the $\sim 3\sigma$ ALMA detection from the ALCS survey \citep{fujimoto2024} as a red contour. The JWST imaging of the CRG clearly reveals a ring morphology; we have scaled the channels of the RGB cutout to emphasize the clumpy blue ring surrounding the red center of the system.}
    \label{fig:mosaic}
\end{figure*}

\section{Observations}\label{sec:obs}

The target of this study is a candidate ring galaxy found on the periphery of the MACS0416 cluster field at $\alpha = 04^{\mathrm h}16^{\mathrm m}11.11^{\mathrm s}$, $\delta = -24^\circ05'24.9''$ and was observed by JWST for the Prime Extragalactic Areas for Reionization and Lensing Science (PEARLS) project \citep[][]{windhorst2023}. The source is cataloged as [MAC2016] M0416cl 183 by \cite{merlin2016} with a photometric redshift $z_{\rm phot} = 3.55 \pm 0.17$, and is cataloged as CANUCS 3100560 in the CAnadian NIRISS Unbiased Cluster Survey \citep{sarrouh2025} with a spectroscopic redshift $z_{\rm spec} = 4.0148$. For simplicity, we refer to this source as ``the CRG'' for the remainder of the work.

We utilize mosaics generated from multi-observatory imaging of this source with the HST ACS instrument \citep{ford2003}, and the JWST NIRCam instrument \citep{rieke2023}. The mosaics were produced as part of the PEARLS project \citep{windhorst2023}, with the HST mosaics produced using the `mosaicdrizzle' pipeline \citep[developed and presented by][] {koekemoer2011}, where the HST data came from several different programs including CLASH \citep{postman2012}, Hubble Frontier Fields \citep[HFF;][]{lotz2017}, BUFFALO \citep{steinhardt2020}, in addition to other public archival data (with the full list of programs being 12459 PI: M. Postman; 13386 PI: S. Rodney; 13459 PI: T. Treu; 13496 PI: J. Lotz; 14209 PI: B. Siana; 15117 PI: C. Steinhardt; 15936 PI: P. Kelly; 15940 PI: B. Ribeiro; 16278 PI: P. Kelly, 16729 PI: P. Kelly, 16757 PI: A. Gonzalez). The HST mosaics were produced from data calibrated with version caldp.20231201 of the HST calibration software\footnote{\url{https://github.com/spacetelescope/hstcal}}, using reference files from HST CRDS\footnote{\url{https://hst-crds.stsci.edu}} context file 1132.pmap.  The JWST mosaics used the JWST pipeline\footnote{\url{https://github.com/spacetelescope/jwst}} version 1.13.4 \citep{bushouse2024}, calibrated with reference files from JWST CRDS\footnote{\url{https://jwst-crds.stsci.edu}} context file 1214.pmap. We make use of the  F606W and F814W bands from HST ACS, and the F090W, F115W, F150W, F182M, F200W, F277W, F356W, and F444W bands from JWST NIRCam. The native flux density units of the mosaics are in nJy, and the pixel size of the mosaics is 0.02$\arcsec$. Further details are presented in the PEARLS overview paper \citep{windhorst2023}. We also make use of a publicly available NIRSpec PRISM spectrum of the source from the CANUCS Data Release 1, as detailed in \cite{sarrouh2025}. 

MACS0416 was also observed as part of an ALMA Large Program (2018.1.00035.L; PI: K. Kohno), the ALMA Lensing Cluster Survey \citep[ALCS;][]{fujimoto2024}, using the Band 6 receiver. Reduced data products for the cluster have been made available on the ALCS website\footnote{\url{https://www.ioa.s.u-tokyo.ac.jp/ALCS/}}. These data products are constructed as a mosaic of the entire cluster; the map uses natural weighting, with a restoring beam size of $1.48\arcsec \times 0.86\arcsec $ and a pixel size of 0.16$\arcsec$. Dust continuum at an observed frequency $\nu_{\rm obs} = 261.3$ GHz is detected at $ \approx 3\sigma$ significance for a flux density rms of 53 $\mu$Jy beam$^{-1}$. After correcting for the beam factor, we obtain a flux density of 240 $\pm$ 80 $\mu$Jy for the dust continuum. 

\begin{figure*}\label{fig:stamps}
    \centering
    \includegraphics[width=0.98\textwidth]{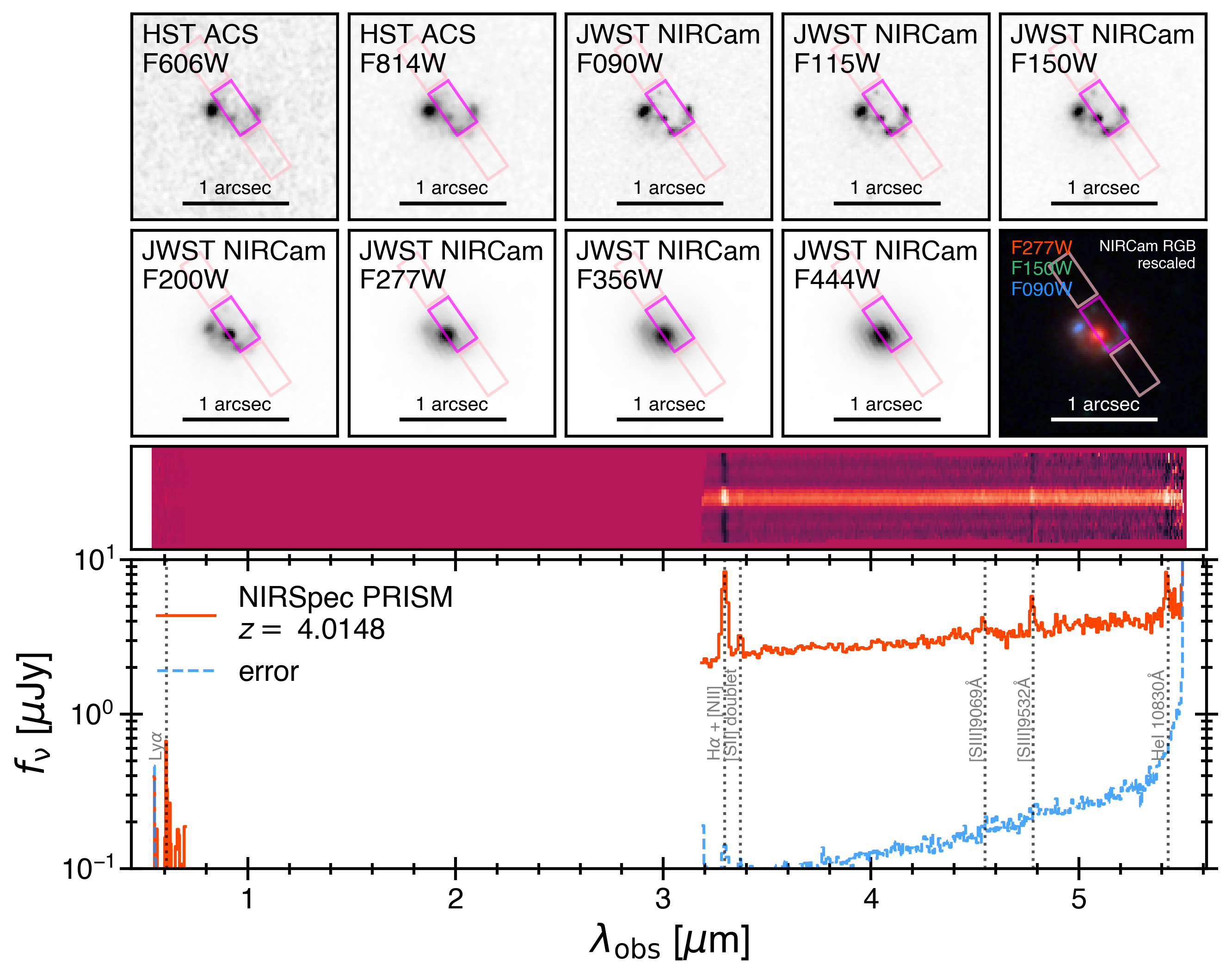}
    \caption{2$\arcsec$ cutouts of the ring galaxy as observed by HST ACS (F606W and F814W) and JWST NIRCam (F090W, F115W, F150W, F200W, F277W, F356W, and F444W) at their native resolutions. The bottom-right cutout panel shows a rescaled RGB image from the NIRCam data using F277W (red), F150W (green), and F090W (blue) to further highlight the red center vs. blue ring. The magenta box shows the region from which the NIRSpec PRISM data were taken. These data products (2D spectrum and integrated 1D spectrum) are shown in the bottom half of the figure. We have scaled the contrast in the cutouts and 2D spectra to highlight important features, showing a red continuum exhibiting strong emission lines from a source at $z = 4.0148$, with emission lines that appear to encompass the spatial extent of the continuum source. The vertical dashed lines indicate the location of expected emission lines at $z = 4.0148$, and are included with annotations to demonstrate a match with the spectroscopic redshift of the galaxy.}
\end{figure*}

\section{Results}\label{sec:res}

In Figure \ref{fig:mosaic} we present a pseudo-RGB view of the cluster field, combining six filters of JWST NIRCam imaging, and two zoomed panels showing the ring galaxy in more detail. The multi-color image of the CRG reveals a red galaxy that is surrounded by a diffuse blue ring, and the high-resolution imaging in F150W clearly shows a ring that is dominated by three bright clumps. We estimate that the ring has a radius of $\approx 0.25\arcsec$ by roughly fitting an ellipse around the ring using the \texttt{CARTA} software \citep{comrie2021}. This distinct combination of color and morphology is why \cite{diego2024} identified this object as a candidate galaxy-galaxy lens. The system also includes some faint features, namely five faint companions within a $\sim 1\arcsec$ radius, and an apparent tidal tail extending from the west end of the galaxy towards the northeast (see Section \ref{sec:ana}). 

The NIRSpec PRISM spectrum covers the center of the galaxy along with at least one clump. We present this spectrum in Figure \ref{fig:stamps}. The published redshift of $z = 4.0148$ from \cite{sarrouh2025} for this source is derived from the PRISM data, with robust detections of the  H$\alpha$ +[\text{N\small II}] complex, the [\text{S\small II}] doublet, the \text{He\small I} line at 1.083 $\mu$m, [\text{S\small III}] lines, and faint Ly$\alpha$. Due to the low spectral resolution of the NIRSpec data, it is challenging to accurately constrain line fluxes, or to deblend some line emission, e.g., the H$\alpha$ flux with the neighboring [\text{N\small II}] lines. The resolution in the NIRSpec data, however, is sufficient to see some structure in the H$\alpha$ +[\text{N\small II}] complex that is consistent with normal star-formation and \text{H\small II} emission. Furthermore, there are no obvious lines that cannot be explained by the spectrum of a star-forming galaxy at $z = 4$. 

The faint ALMA detection of dust continuum, along with its relatively low spatial resolution, makes it difficult to determine whether the bulk of emission arises from the red component or blue ring. We scale the 1.15mm flux density to an 870$\mu$m flux density assuming $\beta = 1.8$, yielding $S_{\rm 870 \mu m} = 0.7 \pm 0.2$ mJy, which we convert into a dust mass using the redshift-independent scaling relation in \cite{dudzeviciute2020} to obtain $M_{\rm dust} = (9 \pm 4) \times 10^7 \ M_{\rm \odot}$. 

\begin{figure*}\label{fig:photometry}
    \centering
    \includegraphics[width=0.98\textwidth]{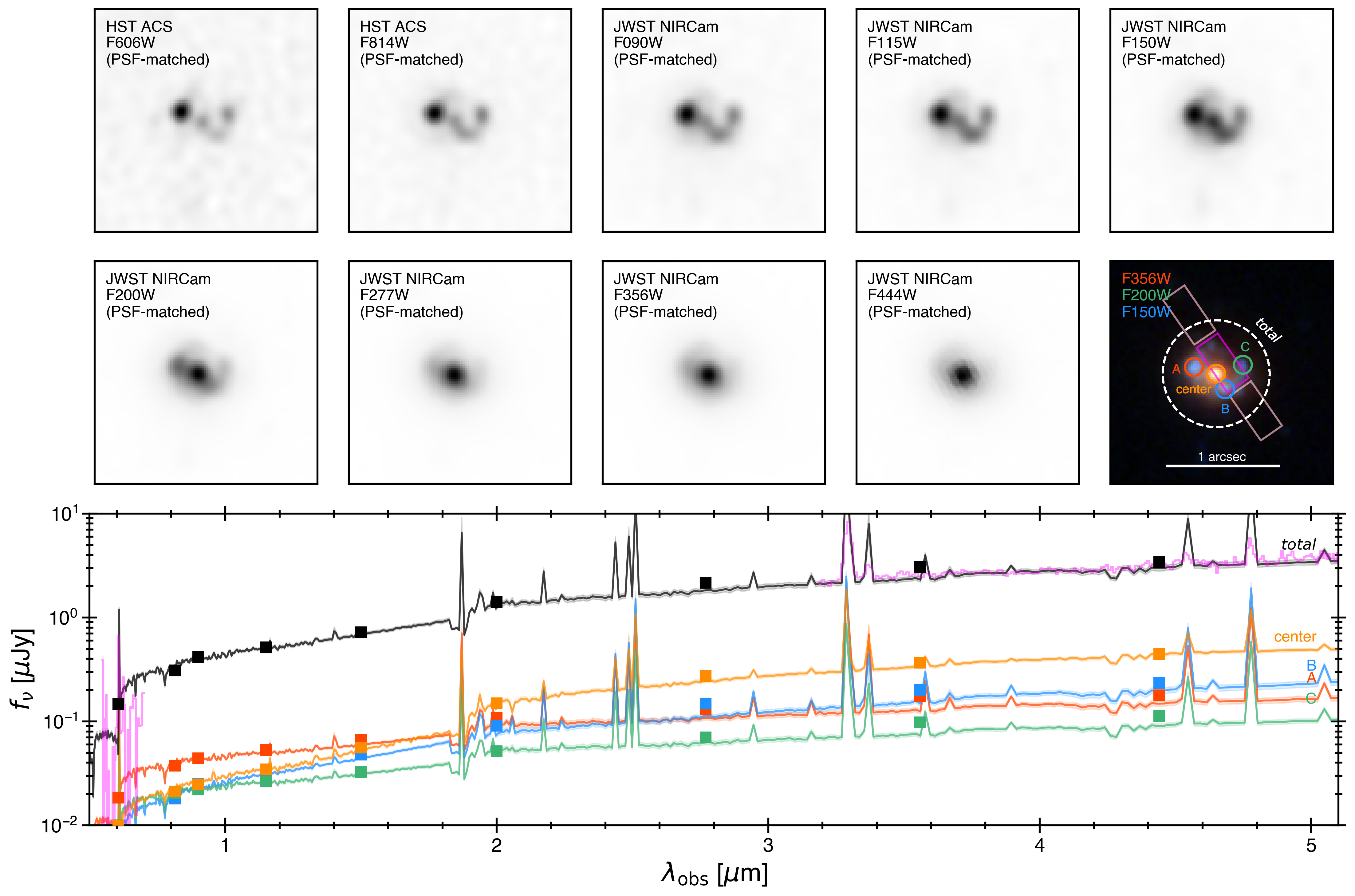}
    \caption{\textbf{Top:} 2$\arcsec$ cutouts of HST ACS and JWST NIRCam imaging, after matching the spatial resolution to F444W. The bottom-right cutout is the RGB image from Figure~\ref{fig:mosaic} at native NIRCam resolution. The scale bar in the RGB cutout applies to all panels. Circles in the RGB panel mark the five apertures used for the multi-band photometry and SED modeling: clumps A (red), B (blue), and C (green), the center (orange) and the total flux (white, dashed line). The magenta rectangle indicates the slit for the spectra in Figure~\ref{fig:stamps}. \textbf{Bottom:} Flux density versus observed wavelength within each aperture, as color coded in the top panel. A 5\% flux uncertainty was multiplied in quadrature to each photometric point before SED fitting. The best-fit SED models from \texttt{bagpipes}, assuming $z = 4.0148$ for all apertures, are shown as solid lines, demonstrating excellent agreement between the photometry and the spectroscopic redshift from CANUCS \citep{sarrouh2025}.
    }
\end{figure*}

There are three scenarios which sufficiently explain the measured spectroscopic redshift for the source. The simplest explanation is that the entire system is a ring galaxy at $z \approx 4$, making it the farthest such system ever detected. This explanation is supported by the 2D PRISM spectrum presented in Figure \ref{fig:stamps}, in which the emission lines peak at the spatial center of the slit. An alternative explanation is that the strong lines detected with NIRSpec are emitted from the ``ring'' and that the red central emission comes from a foreground galaxy; in other words, this is a galaxy-galaxy strong lens system. We discuss these two scenarios in Section \ref{sec:ana}. A third scenario, which we find highly implausible, is that the foreground galaxy is at $z = 4.0148$, and that the blue emission is coming from a background galaxy. Gravitational lenses beyond $z > 2$ are exceedingly rare, even in forecasts of gravitational lens populations in the era of JWST \citep[e.g.][]{hogg2025}. 

\section{Analysis}\label{sec:ana}

\begin{figure*}[ht!]
    \centering
    \includegraphics[width=0.98\textwidth]{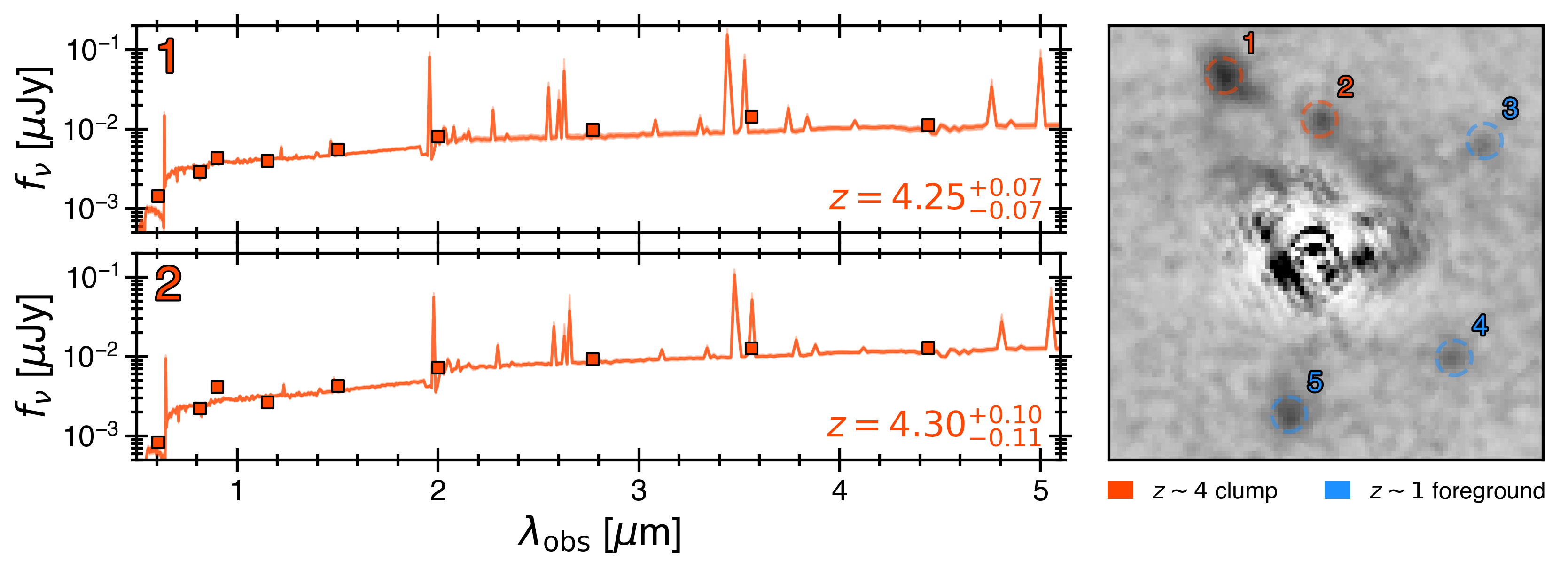}
    \caption{\textbf{Left:} Best-fit SED models (median and 1$\sigma$ error) from photometry of two companions on the outskirts of the CRG with photometric redshifts that are similar to the spectroscopic redshift of the CRG. \textbf{Right:} Model-subtracted, 2\arcsec\ cutout of the F277W image. The subtracted emission is modeled as a 2D S\'ersic profile. The image scaling is tuned to highlight the nearby companions and the arm-like tidal feature extending from the edge of the ring. The locations and aperture sizes used for possible companion galaxies are indicated by the dashed circles. The two clumps with $z_{\rm phot} \sim 4$ are colored red, as in the right panels, while the other clumps, colored in blue, have $z_{\rm phot} \sim 1.0-1.5$, and are most likely foreground sources.}
    \label{fig:outer_companions}
\end{figure*}

To determine whether the CRG is a ring galaxy or a case of gravitational lensing, we employ the \texttt{bagpipes} software \citep{carnall2018} to model the SED of the total galaxy and individual clumps along the ring using multi-band photometry. For this analysis, we utilize the HST ACS F606W and F814W bands, and the JWST NIRCam F090W, F115W, F150W, F200W, F277W, F356W, and F444W bands. The HST ACS PSFs were generated from a publicly available repository of pre-prepared PSF models\footnote{\url{https://astro.dur.ac.uk/~rjm/acs/PSF/}}{\citep[see also][]{rhodes2007}, and resampled from a pixel size of 0.03$\arcsec$ to 0.02$\arcsec$. For the JWST NIRCam PSFs, we use the \texttt{stpsf} \citep[formerly \texttt{WebbPSF;}][]{perrin2014} Python software package to generate simulated point-spread functions (PSFs) for each JWST band, oversampling the PSF to match a target pixel size of 0.02$\arcsec$. Because the F444W band has the lowest spatial resolution, with a FWHM $\sim 0.16\arcsec$, we create a shared kernel between each respective band's PSF via the \texttt{psf.matching} sub-module in \texttt{photutils} \citep{photutils2020} before convolving the F444W-matched kernel with each band of observations. 

We construct four separate apertures for the photometric analysis. One aperture surrounds the red center of the galaxy, and three of them (A, B, C) are centered on bright clumps along the ring. These apertures all have a radius of 0.08$\arcsec$, with the aperture size roughly corresponding to the lowest resolution element. We additionally apply a wide aperture that covers the entire size of the system, covering a radius of 0.48$\arcsec$. 

For this analysis, we employ a delayed star-formation history model in \texttt{bagpipes}, consisting of six parameters: stellar mass $\log_{10}(M_{\rm \star} /M_{\rm \odot})$, star-formation rate (SFR), dust attenuation ($A_V$) as characterized in \cite{calzetti2000}, the system age, metallicity ($Z / Z_{\rm \odot}$), and the ionization parameter ($U$) which controls a simple model of nebular emission. The photometric fitting is performed using nested sampling, specifically using \texttt{nautilus} \citep{lange2023}, as implemented in \texttt{bagpipes}. We additionally apply a systematic error of 5\% in quadrature to the photometric measurements before performing SED modeling. 

We first consider a case in which the redshift is allowed to vary from $0 < z < 5$. In this scenario, we find agreement between the fitted redshifts of the clumps ($3.89 < z < 4.19$), the fitted redshift of the total system ($z_{\rm phot} = 3.92^{+0.10}_{-0.07}$), and the spectroscopic redshift from CANUCS ($z_{\rm spec} = 4.0148$). We emphasize that for this exercise, we have employed a rather simple photometric model, and as such we cannot place strong constraints on certain parameters (e.g., age, metallicity). Even so, this model strengthens the argument for the CRG being a ring galaxy at $z \approx 4$. We then consider a model where the redshift is fixed at $z = 4.0148$. In this case, we find a total star-formation rate $({\rm SFR}) = 140^{+20}_{-30}$ ${\rm M}_{\rm \odot} \ {\rm yr}^{-1}$ and $\log_{10} (M_{\rm \star} / {\rm M}_{\rm \odot}) = 10.41^{+0.11}_{-0.13}$. The results of the SED modeling are shown in Figure \ref{fig:photometry}. 

In total, the three main clumps along the ring have roughly one third the mass of the system's central emission, but have a higher SFR by over a factor of 2. The bulk of the stellar mass and SFR comes from the wider galaxy that is much more prominent in the redder JWST imaging (i.e., redward of F200W). It is also seen in Figure \ref{fig:photometry} that the three clumps have different colors; this is not easily explainable by gravitational lensing without the inclusion of differential dust reddening. While we again emphasize that \texttt{bagpipes} cannot strongly constrain ages, as evidenced by the large uncertainties on this parameter (Appendix Table~\ref{tab:stellar_props_combined}), the central emission is much older ($1.0^{+0.3}_{-0.3}$--$1.2^{+0.2}_{-0.3}$~Gyr for the fitted vs.\ fixed-redshift models, respectively) than the clumps along the ring ($\lesssim$400~Myr with large uncertainties). If the CRG is indeed a collisional ring galaxy, the progenitor galaxy would have had to form around $z \gtrsim 9$ with the ring-forming collision taking place around $z \simeq 5-6$ or later (i.e., $z < 5$).

We apply a similar modeling approach to the five companions on the outskirts of the system, whose locations are shown in Figure \ref{fig:outer_companions}. To figure out where to place apertures, we first remove the foreground light in F277W, using the \texttt{GALFIT} software \citep{peng2002} to fit a single S\`ersic profile with $R_e \sim 0.21\arcsec$, $n \sim 1.2$, and $b/a \sim 0.7$ after masking the faint companions. We then extend the parameter priors in our \texttt{bagpipes} model to allow for low-redshift and high-redshift solutions. Of the five companions visible in Figure \ref{fig:outer_companions}, we find that two have photometric redshifts in agreement with the spectroscopic redshift of the system; both of these lie along an apparent tidal arm visible in F200W and longer wavelength bands. The first companion at the end of this tail has $z_{\rm phot} = 4.24^{+0.07}_{-0.07}$, $\log(M_{\rm \star} / {\rm M}_{\rm \odot}) = 7.74^{+0.14}_{-0.16}$, $A_V = 0.84^{+0.08}_{-0.10}$, and SFR $= 0.5^{+0.0}_{-0.1}$ ${\rm M}_{\rm \odot}$ yr$^{-1}$. The second companion, lying in the middle of the tail, has $z_{\rm phot} = 4.30^{+0.11}_{-0.11}$, $\log(M_{\rm \star} / {\rm M}_{\rm \odot}) = 8.03^{+0.10}_{-0.12}$, $A_V = 0.85^{+0.11}_{-0.10}$, and SFR $= 0.4^{+0.1}_{-0.1}$ ${\rm M}_{\rm \odot}$ yr$^{-1}$. The other three sources have photometric redshifts that are consistent with with $z \simeq 1.0-1.5$. However, to confirm whether any or all of these companions are part of this system, we require high-resolution (spectral and spatial) follow-up observations.

\subsection{Is this system a gravitational lens?}

\begin{figure*}[ht]
    \centering
    \includegraphics[width=0.98\textwidth]{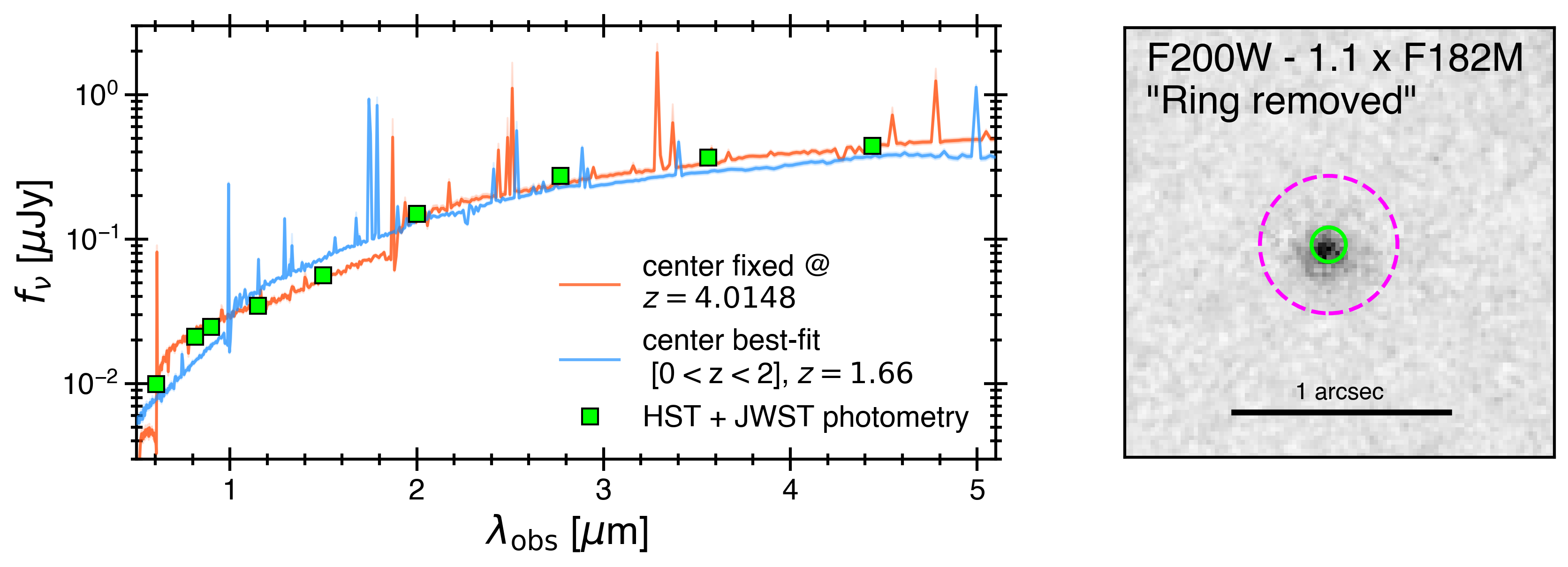}
    \caption{\textbf{Left:} Flux density versus observed wavelength, plotting photometric points from the two HST and seven JWST wide-filter images in the green squares. These data are taken from an aperture corresponding to the ``center'' shown in Figure \ref{fig:photometry} (and the green circle in the right panel). The orange and blue spectra represent \texttt{bagpipes} models of a galaxy clump at $z=4.0148$ versus a foreground gravitational lens with a best-fit redshift of $z = 1.66$. The $z \simeq 4$ solution better describes the photometry, and the predicted source SED at $z \simeq 1.7$ exhibits a number of lines between $3.2 - 5.0$ $\mu$m that are not found in the NIRSpec data (see Figure \ref{fig:stamps}). \textbf{Right:} A scaled version of F182M imaging subtracted from the F200W imaging; we calculate a ``corrective factor'', accounting for diffuse starlight near the ``center'' of the galaxy, from the flux enclosed within the magenta aperture.}
    \label{fig:z_comparison}
\end{figure*}

Until now we have only demonstrated abundant circumstantial evidence pointing to the CRG being a ring galaxy at $z \simeq 4$. However, we do not have the requisite observational data to prove the ring galaxy hypothesis conclusively. Therefore, in this section we consider the second possible scenario, where the red center is an ``early type'' foreground lens at a lower redshift. 

In order to consider the gravitational lensing hypothesis, we first attempt to probe diffuse starlight near the center of the ``lens'' with which we can correct photometry for accurate stellar masses and SFR. This requires us to subtract the ring from the imaging of at least one band. The F182M filter covers the [\text{O\small II}] line at rest-frame 3727Å (assuming emission at $z = 4.0148$), and by assuming that the [\text{O\small II}] emission mostly arises from the ring, and that the spatial distribution of the stars and ionized gas within the ring are morphologically similar, we can attempt to get a view of the central emission only. We multiply the F182M imaging by a factor of 1.1, before subtracting it from the F200W imaging, to create a ring-subtracted image with only central emission plus a diffuse component remaining; the resulting image is shown in the right panel of Figure \ref{fig:z_comparison}. We determine a corrective factor of 1.87 by comparing the total flux in the ring subtracted image within a $0.32\arcsec$-radius aperture to the measured F200W photometry in a $0.08\arcsec$-radius aperture. We lastly apply this factor to all photometric measurements within the central component.

With this ``corrected'' photometry, we then perform an alternate run of the photometric fitting with \texttt{bagpipes}. We consider a foreground source with a redshift between $0.1 < z < 2.0$, and an age between 1 Gyr and 5 Gyr, with other parameters remaining the same. The resulting fit yields $z_{\rm fit} = 1.66 \pm 0.01$; in this scenario, the ``foreground'' galaxy has a $\log(M_{\rm \star} / \rm{M}_{\rm \odot}) = 9.14 \pm 0.02$ and $\rm{SFR} = 3.2^{+0.1}_{-0.2}$ $\rm{M}_{\rm \odot}$ yr$^{-1}$, respectively. We also obtain a dust extinction of $A_V \approx 1.94^{+0.04}_{-0.05}$, which may associate the detection of dust continuum with the lens in a lensing scenario. In Figure \ref{fig:z_comparison} we compare the uncorrected photometric model at $z = 1.66$ to the photometric model determined for the collisional ring galaxy explanation, along with a panel showing the aperture used for ``correcting'' the foreground lens for physical parameter estimation. The former solution generally shows agreement with the photometric data redward of 2 micron, whereas the bluer wavelength bands show much better agreement to the ring galaxy explanation. The former also predicts a strong Pa-$\alpha$ feature at $\lambda_{\rm obs} \simeq 5.0$ $\mu$m which does not appear in the NIRSpec PRISM data.

To further assess the lensing hypothesis, we do not attempt to perform detailed gravitational lens modeling, primarily because a lens modeling failure would not necessarily mean that the system is not a galaxy-galaxy lens. Instead, we make a simple calculation to assess the plausibility of the system as a galaxy-galaxy lens. In the thin lens approximation of gravitational lensing, the projected mass enclosed within an Einstein ring is defined as
\begin{equation}
    M_E = \frac{{\theta_E}^2 c^2}{4 G} \frac{D_S D_L}{D_{SL}}
\end{equation}
defining $\theta_E$ as the Einstein radius, $D_S, D_L$ as the cosmological distances to the background source and foreground lens, respectively, and $D_{SL}$ as the distance between source and lens. A source redshift $z = 4.0148$, lens redshift $z = 1.66$, and Einstein radius $\theta_E = 0.25\arcsec$ implies a projected $M_E = 3.9 \times 10^{10} \ M_{\rm \odot}$. 

Dividing the stellar mass determined above ($M_{\rm \star} =$ 1.4 $\times$ 10$^9$ $M_{\rm \odot}$) by the projected mass implies a stellar-to-total mass ratio $\approx 0.04$. This is a very low fraction; it has been argued that nearby dwarf galaxies (i.e., less massive than the source studied in this work) mostly have systematically large fractions of dark matter across all measured radii \citep[e.g.][]{oh2011}, albeit with lower dark matter to total mass fractions ($\sim 0.7$) than in this case ($\gtrsim 95\%$). Estimating a more conservative Einstein radius of 0.2$\arcsec$ does not solve this problem, yielding a stellar-to-total mass ratio $\approx$ 0.06. In both cases, the stellar-to-total mass ratio is lower than what has been found in similar JWST-selected high-redshift ($z_{\rm lens} > 1$) lenses, with ratios of 0.32 and 0.31 in \cite{shuntov2025} ($z_{\rm lens} \sim 2.0$) and \cite{adams2025} ($z_{\rm lens} = 1.26$), respectively. This problem, overall, weakens the case for the CRG being a gravitational lens. Considering all of the above, we believe that the CRG is more likely a high-redshift ring galaxy (at $z = 4.0148$) than a gravitationally lensed system.

\section{Discussion}\label{sec:disc}

\begin{figure*}[ht!]
    \centering
    \includegraphics[width=0.98\textwidth]{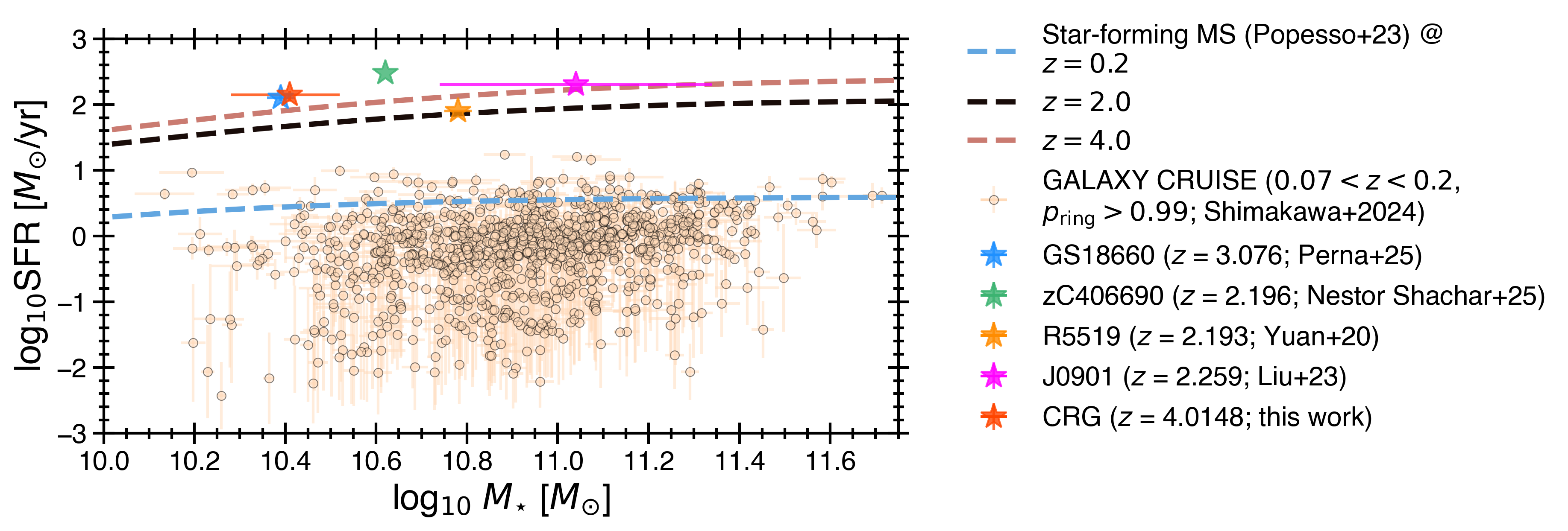}
    \caption{The star-forming main-sequence (SFMS) for a comparison sample of low-redshift ring galaxies \citep{shimakawa2024} and a selected sample of high-redshift ring galaxies \citep{yuan2020, liu2023, nestor2025, perna2025}, including the galaxy studied in this work. We use the parametrized main sequences from \cite{popesso2023} for further comparison. We find that observed high-redshift ring galaxies to date have much higher SFR than their lower-redshift analogs, and lie on or above the SFMS at $z = 2-4$.}
    \label{fig:comparison}
\end{figure*}

\subsection{Comparison to high-redshift ring galaxy candidates}

With the above analysis in mind, we now turn our attention to the small but growing sample of high-redshift ring galaxies \citep{yuan2020, liu2023, nestor2025, li2025, khoram2025, perna2025} for comparison to this work. With a radius of $\sim 0.25\arcsec = 1.8$ kpc, the galaxy studied in this work is smaller in size to those studied in previous works, whose characteristic rings vary in radii from about $3 - 10$ kpc. An important similarity between our work and high-redshift ring galaxies is that these rings are mostly comprised of several star-forming clumps.  \cite{liu2023, nestor2025, perna2025} explicitly characterize their rings as clumpy, and in \cite{li2025}, the ``Cosmic Owl'' comprised of two interacting collisional ring galaxies, a starbursting clump serves as the ``beak''. The fact that the system studied in this work can be characterized as a clumpy ring does not necessarily mean that the ring was formed by a collision; it is well known that clumps generally form due to disk fragmentation driven by gravitational instabilities \citep{toomre1964}, and \cite{dekel2020} argued that most clumpy rings ($z < 4$) do not directly arise from galaxy mergers in high-redshift galaxy simulations \citep[cf.][]{elagali2018}. The clumpy nature of high-redshift rings, nonetheless, is an indicator of gravitational instability along the ring. 

An important difference between the CRG and other high-redshift rings is that the latter generally show clear evidence for a companion/collider galaxy that could have triggered ring formation. The system reported by \cite{liu2023} is notable in that the ring galaxy is gravitationally lensed, making the search for a companion difficult (the ring morphology is determined via forward modeling a gravitational lens model to obtain a source plane morphology). In the case of the CRG, the nearest galaxy in the MACS0416 field with a spectroscopic redshift (CANUCS ID 3100641, $z = 3.9978$) lies $\approx 18\arcsec$ away, corresponding to a physical distance of $\approx$ 125 kpc; as such, it is an unlikely candidate for the dynamical perturber of the ring galaxy. As for the CRG, we have henceforth shown in Section \ref{sec:ana} and Figure \ref{fig:outer_companions} that two faint companions are aligned with a tidal tail that appears to extend from the CRG. As these candidates are low-mass ($\lesssim 10^8$ M$_{\rm \odot}$) compared to the mass of the ring galaxy, it is unlikely that these objects triggered ring formation; a typical assumption for collisional ring galaxy formation is that the ratio of bullet-to-disk mass should be $\sim 0.1$ or greater \citep[e.g.,][]{donghia2008}. Another possible candidate for the galaxy bullet, despite its low mass, is clump ``A'' $(\log (M_{\rm \star} / {\rm M}_{\rm \odot}) = 8.88^{+0.10}_{-0.12})$, whose photometry reveals a starkly different and bluer color compared to the other clumps along the ring. Alternatively, it is possible that line-of-sight effects are hiding the ``bullet'' (e.g., the galaxy is behind the CRG) or that the ``bullet'' was destroyed or dissipated post-collision.

\begin{figure*}[ht!]
    \centering
    \includegraphics[width=0.98\textwidth]{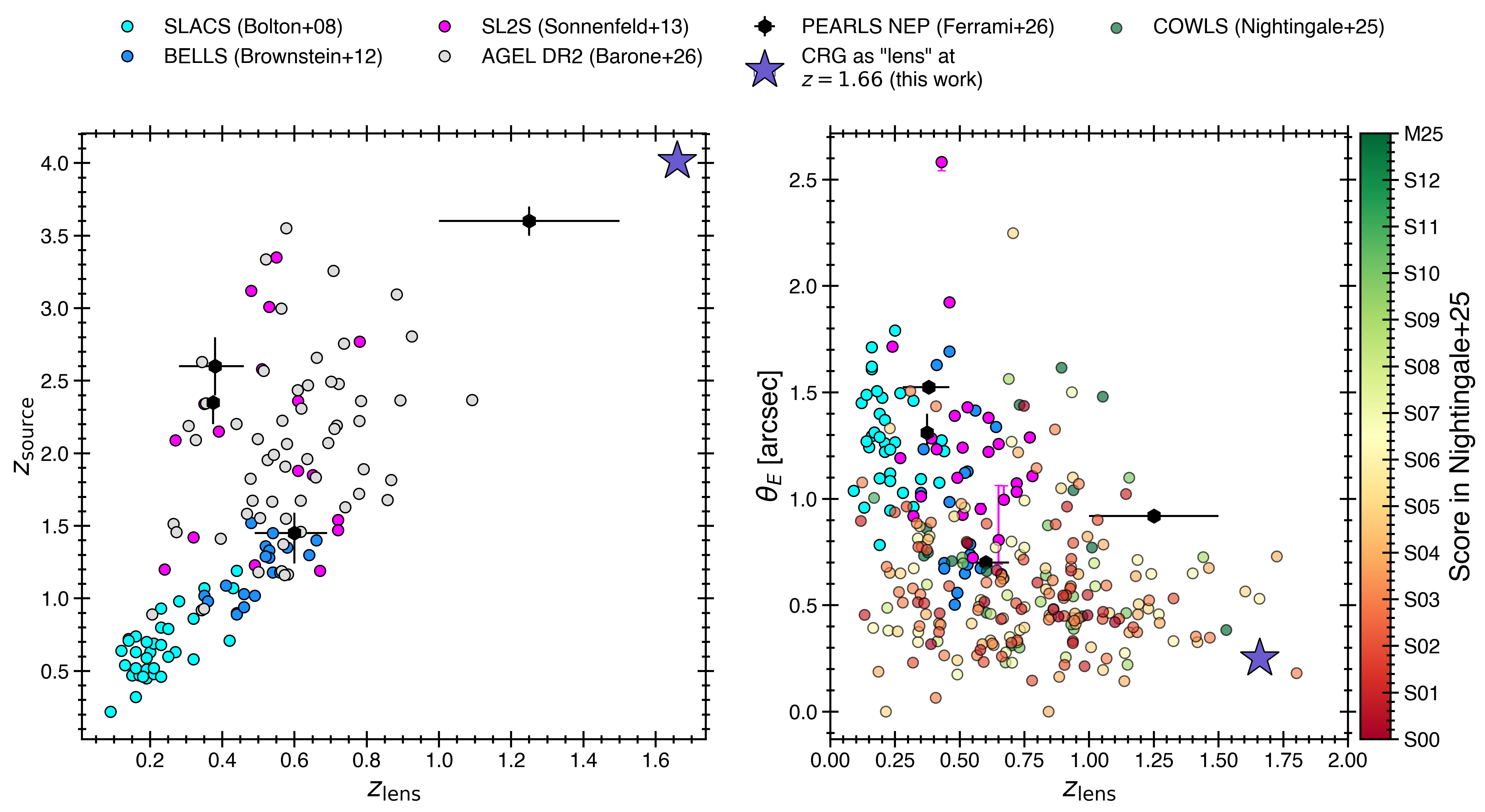}
    \caption{\textbf{Left:} Source versus lens redshift for a sample of well-characterized, optically selected strong lensing systems \citep[SLACS, BELLS, SL2S;][respectively]{bolton2008, brownstein2012, sonnenfeld2013} as compiled in \cite{tan2024}, along with a recent release of strong lenses from the AGEL survey \citep{barone2026} and JWST-selected lenses in the PEARLS NEP time domain field \citep{ferrami2026}. The lensing solution for the CRG is in excellent agreement with the existing parameter space for strong lenses. \textbf{Right:} Einstein radius for SLACS, BELLS, SL2S, PEARLS NEP, and over 400 candidate gravitationally lensed systems in the COWLS program \citep{nightingale2025}. Assuming the CRG is not a lens, it cannot be distinguished from a selection in this parameter space, where it is similar to high-ranked and low-ranked lens candidates.  }
    \label{fig:z_lens_plot}
\end{figure*}

We lastly attempt to compare the modeled physical properties of the CRG to the works cited above, and with a larger sample of local ring galaxies. For the local galaxy sample, we utilize a catalog of ring galaxies from the GALAXY CRUISE survey \citep{shimakawa2024}, placing a cut on redshift between $0.07 < z < 0.2$, SFR $> 10^{-3}$ $\rm{M}_{\rm \odot}$ yr$^{-1}$, and only selecting galaxies with a ring probability $p> 0.99$. We then crossmatch these galaxies to the GALEX-SDSS-WISE Legacy Catalog \citep[GSWLC;][]{salim2016, salim2018} for SFR and stellar masses derived from \texttt{CIGALE} \citep{boquien2019}, returning a final sample of 1,021 local ring galaxies. We plot the SFR versus stellar mass (i.e., the star-forming main sequence) in Figure \ref{fig:comparison}. It is seen that the stellar mass and SFR for our galaxy are similar to those of other high-redshift ring galaxies. We compare both samples to parametrized star-forming main sequences at $z = 0.2, 2.0, 4.0$ using the functional form from \cite{popesso2023}, corresponding to characteristic redshifts for the local and high-redshift samples respectively. In the local sample, many galaxies lie below the star-forming main sequence at $z \approx 0.2$; this is not surprising, as it has been argued in some works \citep[e.g.,][]{kelvin2018, smith2022, fernandez2024} that ring galaxies are overrepresented in the ``green valley'' compared to other morphological classes of galaxies. The high-redshift ring galaxies (including the CRG), conversely, are shown to lie on or above the star-forming main sequence at their respective redshifts, although this is most likely a selection effect.

\subsection{Implications of ring galaxies for future gravitational lens searches}

Finally, we discuss the impact of high-redshift ring galaxies, such as the one studied in this work, as a source of contamination in current and future surveys for gravitational lenses. It is anticipated that $\mathcal{O}(10^5)$ galaxy-galaxy scale gravitational lenses will be discovered in future large sky surveys \citep[e.g.,][]{collett2015}. With the launch of the Euclid mission, early searches for these systems are already underway with what has been called the Strong Lensing Discovery Engine \citep[e.g.,][]{euclid2025}. It is seen in \cite{lines2025} that a number of highly-ranked strong lens candidates, as scored and detected by a fine-tuned pretrained machine learning model, appear to be candidate ring galaxies. While JWST is not optimized, by design, to discover many strong lenses, it has been used to characterize many hundreds of galaxy-galaxy scale gravitational lenses in surveys such as the COSMOS-Web Lens Survey \citep[COWLS; see][]{nightingale2025, mahler2025, hogg2025}.

In Figure \ref{fig:z_lens_plot} we consider how the ``lensing'' scenario compares to samples of gravitational lenses from the literature. For this exercise we consider canonical samples of optical gravitational lenses, such as SLACS \citep{bolton2008}, BELLS \citep{brownstein2012}, and SL2S \citep{sonnenfeld2013}, the second data release from the AGEL survey \citep{barone2026}, and four gravitational lenses from the PEARLS North Ecliptic Pole Time Domain Field \citep{ferrami2026}. The lensing solution for the CRG is perfectly plausible within the parameter space of source versus lens redshift. We also compare our lensing scenario to the sample of over 400 galaxy-galaxy gravitational lens candidates from COWLS, as presented and scored in \cite{nightingale2025}. Here too, the lensing solution is in excellent agreement with both high-ranking and low-ranking gravitational lens candidates from the COWLS survey, which might suggest that some systems in COWLS, like the CRG, are ring galaxies uncovered by JWST.

Compared to existing optical surveys, JWST is able to probe lower mass gravitational lenses at higher redshift, but it does not appear possible to distinguish lenses from non-lenses based on smaller Einstein radii. From all of the above, we infer that galaxies dominated by a ring morphology remain a significant contaminant in searches for gravitational lenses, which poses challenges to current and future optical and near-infrared surveys.

\section{Conclusion}\label{sec:conc}

In this work, we have performed a follow-up investigation of a gravitational lens candidate on the periphery of the MACS0416 cluster field \citep{diego2024}. Our main findings are:

\begin{enumerate}
    
    \item The candidate ring galaxy (CRG) is spectroscopically identified at $z = 4.0148$  via NIRSpec PRISM data from the CANUCS DR1 \citep{sarrouh2025}. The identification is confirmed due to multiple strong emission line features e.g. Ly$\alpha$, H$\alpha$, \text{He\small I }1.083$\mu$m, [\text{S\small II}], and [\text{S\small III}]. The galaxy is characterized by a blue ring surrounding a red center; the ring is comprised of three main clumps and has a radius of $0.25\arcsec$. Additionally, there are five faint companions within  $1\arcsec$ from the center of the galaxy, and an apparent tidal feature that extends from the edge of the ring. There is also a faint, $ \approx 3\sigma$ dust continuum detection of the CRG from archival ALMA observations.

    \item We derive multi-band photometry from two HST ACS wideband filters (F606W and F814W) and seven JWST NIRCam wideband filters (F090W to F444W) for the three main clumps along the ring, the center of the galaxy, and the whole system. By applying a simple star-formation history model from \texttt{bagpipes} to the photometry, we obtain photometric redshifts for the three clumps and total emission that are in good agreement ($z \sim 3.89 - 4.19$) with the spectroscopic redshift from NIRSpec PRISM. When fixing $z = 4.0148$ to the photometric fits, we find a ($\log M_\star / \rm{M}_{\rm \odot}$) $= 10.41^{+0.11}_{-0.13}$ and SFR $= 140^{+20}_{-30}$ $\rm{M}_{\rm \odot}$ yr$^{-1}$ for the CRG.

    \item We cannot conclusively rule out a lensing explanation for the CRG, and consider whether the red central component could be a foreground lensing galaxy. After performing a crude subtraction of the ring using medium-band JWST data, we find that a galaxy SED with $z_{\rm phot} = 1.66 \pm 0.01$, $\log M_\star / \rm{M}_{\rm \odot} = 9.14 \pm 0.02$ and SFR $= 3.2^{+0.1}_{-0.2}$ $\rm{M}_{\rm \odot}$ yr$^{-1}$ achieves a fit to scaled photometry from the central emission in the CRG. Such a galaxy would require a very low stellar mass ratio ($ \approx 5\%$ of the projected mass within the Einstein radius), making it a less plausible explanation.  

    \item Considering the CRG to be a ring galaxy, we find that its estimated stellar mass, SFR from \texttt{bagpipes}, along with the clumpy nature of the ring, are comparable to other high-redshift (collisional) ring galaxies \citep[e.g.][]{yuan2020, liu2023, perna2025, nestor2025, khoram2025} at $z>2$. Unlike other high-redshift ring galaxies, however, the CRG lacks an obvious candidate for the ``bullet'' (i.e., merger progenitor) that triggered ring formation. We identify several ``bullet'' candidates, including a spectroscopically confirmed galaxy over 100 kpc away, two companions lying within the tidal tail, and the brightest companion along the ring. We find that high-redshift ring galaxies detected with JWST generally lie along or above the star-forming main sequence \citep[SFMS; e.g.][]{popesso2023}, compared to a sample of local ring galaxies that mostly lie below the SFMS. 

\end{enumerate}

We conclude that the CRG is most likely a ring galaxy at $z = 4.0148$, specifically a collisional ring formed in the aftermath of a merger event. If this is confirmed with follow-up observations, this system would be the most distant ring galaxy ever reported in the literature. NIRSpec IFU observations, aiming to spatially resolve the H$\alpha$ line across the entire galaxy and its nearby environment, would be the most straightforward way to confirm the nature of the CRG\footnote{The [\text{C\small II}] fine-structure line at $158\mu$m would be a natural line to target in this galaxy using the Atacama Large Millimeter/submillimeter Array (ALMA); however, it would emit at an observed frequency of 379 GHz, which is outside the frequency coverage of the Band 7 ($275-373$ GHz) and Band 8 ($385 - 500$ GHz) receivers on ALMA.}.

As ring galaxies are known to be a contaminant in ongoing strong lensing surveys \citep[e.g.][]{lines2025}, we predict that observatories such as JWST and Euclid will provide statistically significant samples of ring galaxies out to high redshifts. Future studies of unusual galaxy morphologies at high-redshift should consider that some ringed systems may be real ring galaxies, and not strong lenses. Larger, more detailed studies are required to determine the true abundances of these galaxies and how they may contaminate strong lens selections.

\section*{Acknowledgements}

We thank the anonymous referee for their feedback, which strengthened this work. The JWST data is available at MAST: \dataset[doi: 10.17909/qydg-zt87]{\doi{10.17909/qydg-zt87}}. DV thanks Kedar A. Phadke for providing sample scripts that helped enable the photometric analysis in this work. This material is based upon work supported by the National Science Foundation Graduate Research Fellowship under Grant No.\ DGE 21\~46756. This work is based in part on observations made with the NASA/ESA/CSA James Webb Space Telescope. The data were obtained from the Mikulski Archive for Space Telescopes at the Space Telescope Science Institute, which is operated by the Association of Universities for Research in Astronomy, Inc., under NASA contract NAS 5-03127 for JWST. These observations are associated with programs\# 1176 and 1208. This paper makes use of the following ALMA data: ADS/JAO.ALMA \# 2018.1.00035.L. ALMA is a partnership of ESO (representing its member states), NSF (USA), and NINS (Japan), together with NRC (Canada) and NSC and ASIAA (Taiwan) in cooperation with the Republic of Chile. The Joint ALMA Observatory is operated by ESO, AUI/NRAO, and NAOJ. This research has made use of NASA's Astrophysics Data System and Astrophysics Data System Bibliographic Services. This work made use of the MyFilter web tool \citep{myfilter2023} for data analysis. This research has made use of the NASA/IPAC Extragalactic Database, which is funded by the National Aeronautics and Space Administration and operated by the California Institute of Technology. This research has made use of the SIMBAD database, CDS, Strasbourg Astronomical Observatory, France. I.S. acknowledges support from STFC (ST/X001075/1). RAW and SHC acknowledge support from NASA JWST Interdisciplinary Scientist grants NAG5-12460, NNX14AN10G and 80NSSC18K0200 from GSFC. This work is sponsored (in part) by the Chinese Academy of Sciences (CAS) through a grant to the CAS South America Center for Astronomy. This work is supported by the China Manned Space Program with grant no. CMS-CSST-2025-A07. C.C. is supported by Chinese Academy of Sciences South America Center for Astronomy (CASSACA) Key Research Project E52H540301. PSK acknowledges financial support from the Knut and Alice Wallenberg Foundation.

\bibliography{sample631}

\appendix 

Here we include a table of three \texttt{bagpipes} model results, corresponding to three morphological explanations explored in this work. The top row of Table \ref{tab:stellar_props_combined} employs a blind photometric fit to determine whether the redshift of the ring and the central emission is the same. The middle row of Table \ref{tab:stellar_props_combined} contains the parameters used for analyzing the system studied in this work as a (collisional) ring galaxy at $z = 4.0148$. The bottom row of Table \ref{tab:stellar_props_combined} considers the possibility of the central emission arising from a dusty foreground galaxy between $0 < z < 2$. In Table \ref{tab:aperture_photometry} we include the photometry from HST and JWST to which the \texttt{bagpipes} models are fitted.

\begin{table*}[ht]
\centering
\begin{tabular*}{\textwidth}{@{\extracolsep{\fill}}lcccccc}
\hline\hline
Aperture & $z$ & $\log(M_\star/{\rm M}_\odot)$ & SFR [${\rm M}_\odot\,\mathrm{yr}^{-1}$] & $A_V$ [mag] & Age [Myr] & $Z/Z_\odot$ \\
\hline

\multicolumn{7}{c}{$z_{\rm phot}$ as free parameter} \\
\hline
A      & $4.11^{+0.05}_{-0.08}$ & $8.94^{+0.13}_{-0.11}$ & $6.6^{+0.5}_{-0.9}$  & $0.88^{+0.07}_{-0.09}$ & $260^{+180}_{-100}$  & $0.74^{+0.18}_{-0.21}$ \\
B      & $4.19^{+0.06}_{-0.07}$ & $9.19^{+0.12}_{-0.11}$ & $12.7^{+0.9}_{-1.3}$  & $1.44^{+0.07}_{-0.09}$ & $240^{+160}_{-80}$  & $0.81^{+0.13}_{-0.17}$ \\
C      & $3.89^{+0.11}_{-0.07}$ & $8.70^{+0.12}_{-0.10}$ & $3.8^{+0.3}_{-0.6}$  & $0.99^{+0.08}_{-0.11}$ & $260^{+180}_{-90}$  & $0.94^{+0.05}_{-0.09}$ \\
Center & $3.73^{+0.06}_{-0.08}$ & $9.76^{+0.06}_{-0.07}$ & $11.6^{+2.5}_{-1.9}$ & $1.52^{+0.10}_{-0.09}$ & $1000^{+300}_{-300}$ & $0.36^{+0.23}_{-0.17}$ \\
Total  & $3.92^{+0.10}_{-0.07}$ & $10.35^{+0.11}_{-0.11}$ & $146^{+14}_{-25}$ & $1.34^{+0.10}_{-0.12}$ & $320^{+220}_{-110}$  & $0.71^{+0.18}_{-0.21}$ \\

\hline
\multicolumn{7}{c}{$z_{\rm phot} = z_{\rm spec} = 4.0148$} \\
\hline
A      & --- & $8.88^{+0.10}_{-0.12}$ & $6.6^{+0.4}_{-0.5}$ & $0.93^{+0.06}_{-0.06}$ & $210^{+100}_{-80}$ & $0.77^{+0.16}_{-0.19}$ \\
B      & --- & $9.05^{+0.12}_{-0.12}$ & $12.2^{+0.8}_{-1.9}$ & $1.57^{+0.05}_{-0.07}$ & $140^{+90}_{-50}$ & $0.84^{+0.11}_{-0.19}$ \\
C      & --- & $8.79^{+0.13}_{-0.12}$ & $3.4^{+0.5}_{-0.6}$ & $0.88^{+0.08}_{-0.10}$ & $390^{+270}_{-150}$ & $0.95^{+0.03}_{-0.07}$ \\
Center & --- & $9.86^{+0.04}_{-0.05}$ & $11^{+2}_{-2}$ & $1.36^{+0.09}_{-0.11}$ & $1200^{+200}_{-300}$ & $0.24^{+0.18}_{-0.08}$ \\
Total  & --- & $10.41^{+0.11}_{-0.13}$ & $140^{+20}_{-30}$ & $1.28^{+0.08}_{-0.10}$ & $390^{+240}_{-150}$ & $0.80^{+0.13}_{-0.21}$ \\

\hline
\multicolumn{7}{c}{Central emission as speculated foreground lens} \\
\hline
Center & $1.66^{+0.01}_{-0.01}$ & $9.14^{+0.02}_{-0.02}$ & $3.2^{+0.1}_{-0.2}$ & $1.94^{+0.04}_{-0.05}$ & $1030^{+50}_{-20}$ & $0.99^{+0.01}_{-0.02}$ \\

\hline
\end{tabular*}

\caption{Results from \texttt{bagpipes} \citep{carnall2018} photometric fitting in five apertures applied to the CRG. We target the three main star-forming clumps (A, B, C) along the ring, the center of the object, and the total emission.
\textbf{Top}: photometric redshift is allowed to vary freely. 
\textbf{Middle}: fixed redshift $z=4.0148$ from spectroscopy. 
\textbf{Bottom}: central emission rescaled to account for faint flux, where photometric redshift is allowed to vary between $0 < z < 2$.}
\label{tab:stellar_props_combined}
\end{table*}

\begin{table*}[ht]
\centering
\begin{tabular*}{\textwidth}{@{\extracolsep{\fill}}lccccc}
\hline\hline
Band & A [nJy] & B [nJy] & C [nJy] & Center [nJy] & Total [nJy] \\
\hline
F606W & $18.50 \pm 0.08$ & $6.64 \pm 0.08$ & $10.01 \pm 0.08$ & $9.91 \pm 0.08$ & $147.0 \pm 0.4$ \\
F814W & $37.41 \pm 0.08$ & $18.02 \pm 0.07$ & $20.36 \pm 0.07$ & $21.09 \pm 0.07$ & $309.0 \pm 0.4$ \\
F090W & $44.21 \pm 0.07$ & $25.12 \pm 0.06$ & $22.18 \pm 0.06$ & $24.77 \pm 0.06$ & $418.6 \pm 0.3$ \\
F115W & $53.06 \pm 0.07$ & $34.27 \pm 0.06$ & $26.51 \pm 0.06$ & $34.61 \pm 0.06$ & $514.7 \pm 0.3$ \\
F150W & $66.07 \pm 0.07$ & $47.88 \pm 0.06$ & $32.38 \pm 0.06$ & $56.01 \pm 0.06$ & $721.6 \pm 0.3$ \\
F200W & $108.67 \pm 0.07$ & $90.78 \pm 0.07$ & $51.57 \pm 0.06$ & $149.7 \pm 0.08$ & $1402.6 \pm 0.3$ \\
F277W & $130.23 \pm 0.09$ & $148.73 \pm 0.09$ & $70.28 \pm 0.07$ & $274.05 \pm 0.12$ & $2160.0 \pm 0.4$ \\
F356W & $175.76 \pm 0.11$ & $201.35 \pm 0.12$ & $97.86 \pm 0.09$ & $365.90 \pm 0.15$ & $3045.7 \pm 0.5$ \\
F444W & $177.8 \pm 0.7$ & $233.8 \pm 0.8$ & $113.0 \pm 0.6$ & $443.5 \pm 1.0$ & $3418 \pm 4$ \\
\hline
\end{tabular*}
\caption{Aperture photometry for the galaxy analyzed in this work, measured in two \textit{HST} and seven \textit{JWST} filters. As the error on the bulk of the photometry is very small ($< 1$ nJy), we apply a systematic uncertainty of 5\% to the photometric measurements in \texttt{bagpipes}.}
\label{tab:aperture_photometry}
\end{table*}

\end{document}